\documentclass{aastex}
\usepackage{epsfig}
\usepackage{emulateapj5}

\slugcomment{June 12, 2002 - accepted to Ap.J.Letter}

\def\ea{{\it et al.\,}}
\def\be{\begin{equation}}
\def\ee{\end{equation}}

\def\ni{\noindent}

\begin{document}

\title{MITO Measurements of the Sunyaev-Zeldovich Effect\\ in the Coma
Cluster of Galaxies}

\author{Marco De Petris\altaffilmark{1}, Livia D'Alba\altaffilmark{1},
Luca Lamagna\altaffilmark{1}, Francesco
Melchiorri\altaffilmark{1}, Angiola Orlando\altaffilmark{1},
Emilia Palladino\altaffilmark{1}, Yoel Rephaeli\altaffilmark{2},
Sergio Colafrancesco\altaffilmark{3}, Ernst
Kreysa\altaffilmark{4}, and Monique Signore\altaffilmark{5}}



\begin{abstract}

We have measured the Sunyaev-Zeldovich effect towards the Coma
cluster (A1656) with the MITO experiment, a 2.6-m telescope
equipped with a 4-channel 17 arcminute (FWHM) photometer.
Measurements at frequency bands 143$\pm$15, 214$\pm$15, 272$\pm$16
and 353$\pm$13 GHz, were made during 120 drift scans of Coma. We
describe the observations and data analysis that involved
extraction of the SZ signal by employing a spatial and spectral
de-correlation scheme to remove a dominant atmospheric component.
The deduced values of the thermal SZ effect in the first three
bands are $\Delta T_{0}$=$-$179$\pm$38,$-$33$\pm$81,170$\pm$35
$\mu$K in the cluster center. The corresponding optical depth,
$\tau$=(4.1$\pm$0.9)$\times 10^{-3}$, is consistent (within
errors) with both the value from a previous low frequency SZ
measurement, and the value predicted from the X-ray deduced gas
parameters.
\end{abstract}

\keywords{cosmology: cosmic microwave background -- observations
-- galaxies: clusters: individual (A1656)}


\begin{deluxetable}{rrrrrrrr}
\footnotesize \tablewidth{0pt} \tablecaption{(see text for
explanation)\label{tab:ratios}} \tablecolumns{8} \tablehead{
\colhead{Channel} & \colhead{$\nu_i$} & \colhead{$\langle a
\rangle_i$} & \colhead{$\langle a \rangle_i$} & \colhead{$b_i$} &
\colhead{$c_i$}  & \colhead{$A\Omega_i$} & \colhead{$R_i$}\\
\colhead{$i$} & \colhead{(GHz)} &  \colhead{2000} & \colhead{2001}
& \colhead{} & \colhead{} & \colhead{(cm$^2$ sr)} & \colhead{($\mu
\rm{K}/n \rm{V}$)} }

\startdata 1 & 143$\pm$15 & $1$ & $1$ & $1$ & $1$ & 0.40$\pm$0.02
& 433$\pm$40\\ 2 & 214$\pm$15 & 1.51$\pm$0.45 & 2.59$\pm$1.54 &
$0.11^ {+0.01}_{-0.01}$ & $1.18 ^{+0.00}_{ -0.07}$ & 0.36$\pm$0.02
& 371$\pm$35\\ 3 & 272$\pm$16 & 2.96$\pm$1.21 & 4.96$\pm$2.75 &
$-0.93 ^ {+0.11}_{-0.00}$ & $1.08 ^{+ 0.01}_{-0.08}$ &
0.34$\pm$0.02 & 400$\pm$40\\ 4 & 353$\pm$13 & 12.42$\pm$2.60 &
15.98$\pm$8.76 & $-1.45^{+0.44}_{-0.24}$ & $0.89 ^{+0.12}_{-0.16}$
& 0.34$\pm$0.02 & 347$\pm$40\\

\enddata
\end{deluxetable}

\section{INTRODUCTION}


Compton scattering of the cosmic microwave background (CMB)
radiation by electrons in the hot gas in clusters of galaxies --
the Sunyaev-Zeldovich (SZ) effect -- has long been recognized as a
uniquely important feature, rich in cosmological and astrophysical
information (Zeldovich \& Sunyaev 1969; Sunyaev \& Zeldovich 1970;
Boynton \& Melchiorri 1978; Gunn 1978; Silk \& White 1978;
Cavaliere et al. 1979).

Interest in the effect heightened when high quality
\\
\smallskip
\\
--------------------------------------------------------------------------- \\
{\footnotesize{$^1$Department of Physics, University La Sapienza,
Rome, Italy
\\ marco.depetris@roma1.infn.it}
\\
\footnotesize{$^2$School of Physics and Astronomy, Tel Aviv
University, and Center for Astrophysics and Space Sciences,
University of California, San Diego - yoelr@mamacass.ucsd.edu}
\\
\footnotesize{$^3$Osservatorio Astronomico di Monte Porzio, Rome,
Italy \\ cola@coma.mporzio.astro.it}
\\
\footnotesize{$^4$Max Planck Institute fur Radioastronomie, Bonn,
Germany \\ p464erk@mpifr-bonn.mpg.de}
\\
\footnotesize{$^5$LERMA, Observatoire de Paris, Paris, France \\
monique.signore@mesiob.obspm.fr}}
\bigskip
\\
images of the effect were obtained with interferometric arrays
(Carlstrom et al. 2001; for general reviews, see Rephaeli 1995a,
Birkinshaw 1999).

Accurate determinations of the Hubble constant ($H_0$), cluster
mass profiles, and cluster peculiar velocities from SZ and X-ray
measurements require precise description of the intracluster (IC)
gas temperature and density profiles, and improved control of
systematic errors. These can be {\it more optimally} achieved by
multi-frequency SZ measurements, and high-quality spectral and
spatial X-ray measurements of {\it nearby} clusters.

Therefore, it is essential to develop sensitive SZ experiments
operating at telescopes with effective beam sizes of a few
arcminutes, equipped with bolometer arrays operating at
frequencies on both sides of the Planckian peak. One such
ground-based system is MITO (Millimetre and Infrared Testagrigia
Observatory; De Petris et al. 1996).

Coma is the richest nearby cluster whose IC gas properties have
been determined from extensive X-ray observations. Attempts to
measure the S-Z effect in Coma at low frequencies were reported by
Pariskii (1973), Rudnick (1978), Lake \& Partridge (1980),
Birkinshaw et al. (1981), and Herbig et al. (1995, revised in
Mason et al. (2001)) while at high frequencies by Silverberg et
al. (1997).

\section{OBSERVATIONS}

\subsection{MITO}

MITO is a 2.6-m alt-azimuthal telescope located at an altitude of
3480 m in the Italian Alps. The telescope design, optical
configuration, and electro-mechanical modulation system are
described in De Petris et al. 1989, Gervasi et al. 1998, Mainella
et al. 1996. The 4-channel photometer (Orlando et al. 2002)
consists of 4 NTD Ge composite bolometers cooled down to 290 mK.
The thermodynamic NET is around 1 mK s$^{1/2}$. At the telescope
focal plane the photometer has a 17$'$ FWHM field of view (FOV).

Observations were made during February and March of 2000 \& 2001,
when data from some 120 drift scans of Coma were collected. Each
scan consists of 10 minutes of measurements collected every
second. A 41$'$ beamthrow was selected for a 3-field square-wave
like spatial modulation. The position of Coma in the sky covered
altitude values from 50 up to 73 degrees above the horizon. The
drift scan systematics are better controlled than in source
tracking mode, mainly stemming from weaker microphonics and lower
changes in sidelobe spillover effects. At a data sampling rate of
1 Hz, pointing information, signals, modulation system and
photometer performance were recorded. A pointing accuracy of $\sim
1'$ was attained by frequent measurements of bright stars close to
the source by a CCD camera. We measured beam shape, responsivity
and atmospheric transmission for each channel by means of
planetary calibrators -- Jupiter, Saturn and the Moon -- before
each Coma observation session. In order to correct for atmospheric
absorption we measured the channels zenith transmission during
each night.

\subsection{Results from X-ray Observations}

Coma has been extensively observed, mostly in the 1-10 keV range.
The most recent high spectral and spatial resolution measurements
with the XMM/EPIC experiment are best fit by $kT$=8.2$\pm$0.1
$keV$ (at 90\% confidence level) in the central 10$'$ region
(Arnaud et al. 2001). Analysis of the emission from a larger
region of the cluster yields $kT$=8.2$\pm$0.4 $keV$, which we
adopt here. This value of the temperature is in very good
agreement with the value deduced from Ginga observations. Since
quantitative results for the gas density profile from neither XMM
nor {\it Chandra} are currently available, we use the ROSAT
(Briel, Henry, \& Bohringer 1992) deduced values $n_{e0} \simeq
(2.89 \pm 0.04)\times 10^{-3}$ cm$^{-3}$, $r_c \simeq 10.5' \pm
0.6'$, and $\beta \simeq 0.75$, for the central electron density,
core radius, and index in the expression for the commonly used
$\beta$ density profile, $n_{e}(r) =
n_{e0}(1+r^{2}/r_{c}^{2})^{-3\beta/2}$.

\section{Data Analysis}

Clearly, there is no unique approach to the analysis of a noisy
dataset which includes a dominant, fluctuating atmospheric
emission, and other confusing signals. When the spatial
distribution of the confusing source is unknown, such as that of
CMB anisotropy, a sufficient number of channels is selected so as
to allow removal of anticipated foregrounds but the corresponding
system of equations can often be numerically unstable.

Our task is made feasible largely by the known position of the
source, and its approximate angular extent. Therefore, to first
approximation, we can fit the data with a signal of a known shape
but unknown peak amplitude. The errors in measuring this amplitude
approach the intrinsic detector noise if we can successfully
remove the contributions of atmospheric fluctuations and CMB
anisotropy whose spectral shape is known. In addition to the SZ
effect, these signals were modeled separately from all others
which were collectively treated as unidentified noise. While three
channels are in principle sufficient to achieve this goal, we
added a fourth channel centered on the crossover frequency (where
the thermal SZ effect vanishes), in order to check the separation
between CMB and atmospheric emission, and to {\it possibly} obtain
a rough estimate on the value of the kinematic SZ effect.

We have computed the expected spectral ratios of the SZ signals in
channels 2, 3, and 4 with respect to the first channel. The
calculation is relativistically exact and is based on the
treatment of Rephaeli (1995b; see also Rephaeli \& Yankovitch
1997), taking the above quoted range of the gas temperature. At
each frequency we calculated the intensity change due to the
thermal SZ effect, $\Delta I_{sz}$, and convolved it with the
spectral response of the photometer, $\epsilon_{i}$, and the
atmospheric transmittance $\epsilon_{atm}$, which was determined
from a model (Liebe 1985). The expected ratio of the $i$-th to the
first channel is thus,

\be b_i=\frac{A\Omega_i R_i}{A\Omega_1 R_1}
\frac{\int_{0}^{\infty} \Delta I_{sz}(\nu)
\epsilon_i(\nu)\epsilon_{atm}(\nu)d\nu}{\int_{0}^{\infty} \Delta
I_{sz}(\nu)\epsilon_{1}(\nu) \epsilon_{atm}(\nu)d\nu}, \ee where
$R_i$ is the responsivity in the $i$-th channel as measured with
sky calibrators and $A\Omega_i$ is the throughput, which slightly
changed from channel to channel due to the different optical paths
through the telescope and photometer. Uncertainties in the
evaluation of the $b_i$ are due to the relatively small error in
the value of $kT$ and the fluctuating values of atmospheric
transmittance (0.5-2.0 mm of precipitable water vapour).
Similarly, we have calculated the ratios, $c_i$, of the intensity
change due to the primary CMB anisotropy in the four bands. In
Table~\ref{tab:ratios} all the cited ratios are listed with $R_i$
and $A\Omega_i$.


The recorded data in the four channels (i=1,..,4) for each scan
(j=1,..,120) can be written as follows:
\begin{eqnarray}
\Delta S_i^j &=&  a_i^j \Delta V_{{atm}}^j (t)+ b_i \Delta
V_{{sz}}^j (\alpha,\delta) \nonumber
\\ & +& c_i \Delta V_{{cmb}}^j (\alpha,\delta) + \Delta
V_{{offset},i}^j (\alpha,\delta;t) \nonumber \\ &+& \delta
V_{{spike},i}^j (t) + \Delta V^j_{{noise},i} (t).
\end{eqnarray}
In this equation, $b_i \Delta V_{{sz}}^j$ is the SZ signal which
is fit with $w_i^j \Delta V_{{sz-sim}}^j$, where $\Delta
V_{{sz-sim}}^j$ is the simulated SZ signal, normalized to 1 at the
peak; $w_i^j$ are the estimated peak values of SZ signals (in the
absence of noise $\langle w_1 \rangle =b_i^{-1}\langle w_i \rangle
, i=2,3,4$) and $a_i^j$ are the atmospheric ratios with respect to
channel 1, to be determined for each scan. The $a_i^j$ values
could change from scan to scan due to the atmospheric long time
variations while they are assumed to be constant within a single
scan. Since atmospheric fluctuations dominate on each scan,a rough
estimate of the $a_i^j$ can be done as the ratios of the channels
standard deviations. For this purpose we used only the data
outside the expected position of Coma. In Table~\ref{tab:ratios}
we have also quoted the mean values of $a_i$ averaged over all the
drift scans in each observational campaign. Detector and other
sources of noise are collectively included in the term $\Delta
V^j_{{noise},i}$. Estimated errors in the evaluation of $w_i^j$
due to uncertainties in $b_i$ constitute a small fraction of the
overall error.

The data include three main sources of contamination:

\ni {\bf a.} Cosmic ray spikes ($\delta V_{{spike},i}^j(t)$): we
observed roughly one spike in every 10 scans. A special
de-glitching algorithm has been applied to correct these data.
However, their total removal from the analysis did not change our
results apart from a slight increase of the noise. After removal
of the spikes the data were averaged over 15 seconds so that each
drift scan consists of 37 bins roughly separated in sky by 3.75
arcminutes, about $1/4$ of the FOV.

\ni {\bf b.} An instrumental offset (${\Delta V_{{offset},i}^j
(\alpha,\delta;t)}$) ranging from 0.4 $\mu$V (channel 1) up to
5$\mu$V (channel 4) was measured. This is partly due to the
residual side lobes of the telescope, and partly to the thermal
imbalance of the primary mirror. The telescope is stopped during
each drift scan. In this way the offset is obviously independent
of position, so it is possible to remove it from each scan through
a linear fit of the data; a quadratic fit did not change the rms
value of the residual fluctuations. Note that the removal of the
offset through a linear fit may introduce an additional effect,
due to the presence of large atmospheric fluctuations. The average
of these along a finite drift scan will introduce an additional
offset, $\Delta V_{{off-atm},i}^j$ = ${1 \over 37} \sum_{k=1}^{37}
a_i^j \Delta V_{{atm},k}^j (t)$. This unknown offset will in turn
introduce an error in the estimated value of the SZ signal; we
discuss below how this offset was removed.

\ni {\bf c.} Large atmospheric fluctuations; we based our approach
to minimize their effect on the measured data in channel 4, based
on the fact that $a_4$ is $\sim 13$. The efficiency of this
procedure depends on the amplitude of the atmospheric fluctuations
in each drift scan, as well as on the degree of correlation among
the various channels with channel 4. A detailed discussion of the
atmospheric noise will be given elsewhere; here we briefly discuss
only the essential points. In channel 2 the residuals of the
signal after subtracting $a_2^j/a_4^j \Delta S_4^j$ and minimizing
the difference consist of uncorrected atmospheric fluctuations,
signal due to CMB anisotropy, and only a very small SZ signal. If
we fit the residuals with a simulated SZ signal of amplitude
$w_2^j$, we get $\langle w_2 \rangle  \simeq -0.25\pm 0.32$ nV.
This result indicates that the SZ signal in channel 2 is
consistent with zero, as expected, and the large dispersion --
with respect to the final error due to the intrinsic detector
noise of about 0.05 nV -- around the mean suggests that there is a
residual atmospheric contamination $\Delta w_2= a_2 \Delta w_1$. A
significant decrease in the dispersion is obtained if we remove
the values of $w_2$ far from the mean. These values are possibly
due to the presence of a large atmospheric disturbance along the
line of sight (los) to Coma. For instance, if we remove values
which are more than 4$\sigma$ (standard deviation) from the mean,
we obtain $\langle w_2 ^* \rangle \simeq -0.04\pm 0.09$ nV. A
similar analysis has been done for channels 1 and 3, for which we
expect SZ signals at the predicted ratio $b_3$. Residual
atmospheric fluctuations -- as well as the spurious offset
discussed previously -- introduce an additional error. The
corresponding small corrections $\Delta w_1$ and $\Delta w_3= a_3
\Delta w_1$, which have to be added to the measured values of the
SZ effect in order to get the correct value, must satisfy the
relation: $w_1+\Delta w_1 = (w_3+ a_3 \Delta w_1)/b_3$. The
dispersion of ($w_1 + \Delta w_1$) is minimized by changing
$b_3^*$ within $\pm 1.5$ $b_3^*$. The dispersion decreases by a
factor of four when $b_3^*$ coincides with $b_3$ within the
errors.

After correcting for binning and atmospheric transmittance we
obtain $\langle w_1 ^* \rangle  \simeq -0.171 \pm 0.036$, $\langle
w_2 ^* \rangle  \simeq -0.037 \pm 0.090$, and $\langle w_3 ^*
\rangle \simeq 0.174 \pm 0.036$ nV. In Figure 1, we show plots of
$\Delta V_{sz}$, $ \Delta V_{cmb}$, and $ \Delta V_{atm}$ as
derived by solving eq.(2). The data have been filtered with an
adaptive filter which removes spatial patterns that are not
present in the expected SZ profile. The filter does not alter the
SZ signal, but strongly reduces CMB and atmospheric fluctuations.

\smallskip \centerline{\vbox{\epsfxsize=7.5cm\epsfbox{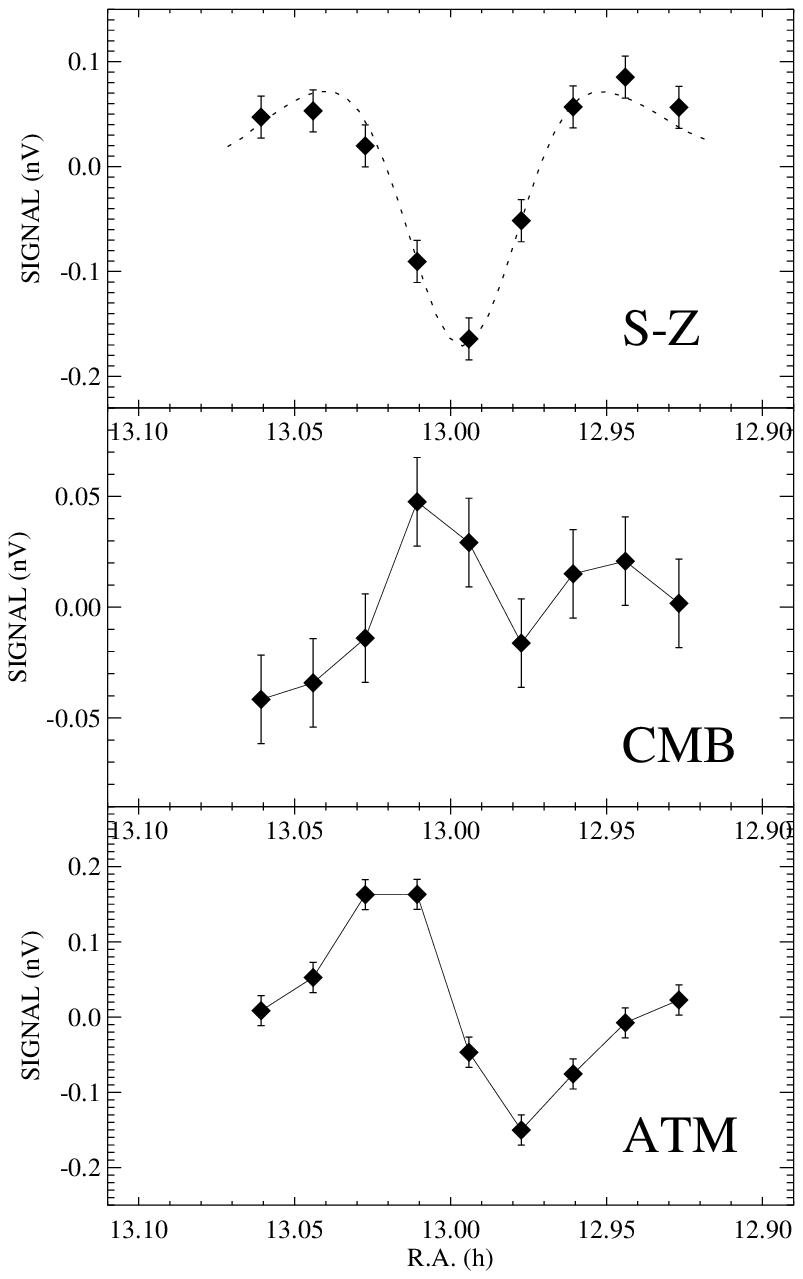}}}
{\ {\small F{\scriptsize IG}.~1. SZ, CMB, and atmospheric signals
extracted from equation (2), corresponding to the 143 GHz band,
averaged over the drift scans and binned in 9 independent sky
positions. Atmospheric fluctuations have an amplitude comparable
to that of the SZ effect, even after 20 hours of integration. CMB
anisotropies are averaged over various parallactic angles
corresponding to the different drift scans.\label{fig:rotor}}}
\medskip

The $w$ value has been converted to thermodynamic temperature for
the $i$-th channel by calibration based on measurements of Jupiter
\cite{Therese}. Integrating its spectral temperature distribution
over the four channels we determined the optical responsivities.
The overall calibration error is about 10$\%$, dominated by the
uncertainty in the brightness temperature of Jupiter ($\sim5\%$),
and statistical errors in the estimate of the beamsize. The final
values of the deduced temperature changes are $\Delta T \simeq$
-73.5$\pm$15.5,-13.7$\pm$33.3,69.6$\pm$14.4 $\mu$K, in the first
three channels, respectively. Based on our calculated value for
the geometrical form factor due to beam dilution and modulation
strategy, $\eta$ = 0.41$\pm$0.02, we compute the corresponding
values $\Delta T_0 \simeq$ -179.3$\pm$
37.8,-33.4$\pm$81.2,169.8$\pm$35.1 $\mu$K for $\Delta T$ in the
cluster center.

\section{DISCUSSION}

MITO measurements of the Coma are the first successful detection
of the SZ effect in this cluster at high frequencies. Our results
imply a mean Thomson optical depth $\tau_0= (4.1\pm 0.9) \times
10^{-3}$ for a los through the center of Coma. Herbig et al.
(1995) have already detected the effect at the much lower
frequency of 32 GHz. Their reported central value, $\Delta
T_0$=$-$505$\pm$92 $\mu$K, and corresponding optical depth,
$\tau_0=(5.6 \pm 1.1)\times 10^{-3}$, were recently updated (Mason
{\it et al.}, 2001) with new calibration data, yielding an
observed decrement $\Delta T$=$-$302$\pm$54 $\mu$K. This value,
corrected for the dilution over their $7.3'$ FOV and the switching
amplitude of $22.2'$, corresponds to $\Delta T_0$=$-$520$\pm$93
$\mu$K and $\tau_0 = (4.0 \pm 0.6)\times 10^{-3}$, and is clearly
in good agreement with X--ray deduced gas parameters. Clearly,
these two results for $\tau_0$ are quite consistent, given the
appreciable ($1\sigma$) errors. Even though our somewhat lower
value is not significantly different from that of Herbig et al.,
it is interesting to note that the CMB anisotropy signal that we
seem to have identified along a los to Coma (heavily exploiting
our multi-frequency capability) could have possibly contributed to
the SZ signal at the low (single) frequency OVRO measurement
(where the two signals cannot be spectrally separated).
Irrespective of whether this is indeed the case, the great
advantage of multi-frequency work has been clearly demonstrated in
the analysis of our measurements.

In Figure 2 we show the first SZ spectrum of Coma by combining
OVRO with our measurements; a fit to the predicted spectrum yields
$\tau_{0} \simeq (4.9 \pm 0.7) \times 10^{-3}$.

In addition to measurement errors, other errors include the use of
a simple one dimensional, isothermal gas density with a $\beta$
profile. Based on the X-ray morphology of Coma, the relatively
small degree of ellipticity of X-ray morphologies of other rich
clusters, and the recent verification of the gas isothermality in
Coma from XMM measurements (Arnaud et al. 2001), we estimate that
the additional error introduced by taking the gas to be
spherically symmetric and isothermal is $\leq 30\%$ of the overall
quoted error. Note also that the additional Comptonization by
non-thermal electrons, (whose existence in Coma is deduced from
measurements of extended radio emission) is negligible (Shimon \&
Rephaeli 2002, Colafrancesco, Marchegiani \& Palladino 2001) if
the energy density in these electrons is at the level deduced from
measurements of high energy ($
> 20$ keV) X-ray emission with the RXTE (Rephaeli, Gruber \&
Blanco 1999) and BeppoSAX (Fusco-Femiano et al. 1999) satellites.
This result is valid also for other {\it viable} energetic
electron models (Shimon \& Rephaeli 2002).

\medskip
\centerline{\vbox{\epsfxsize=7.2cm\epsfbox{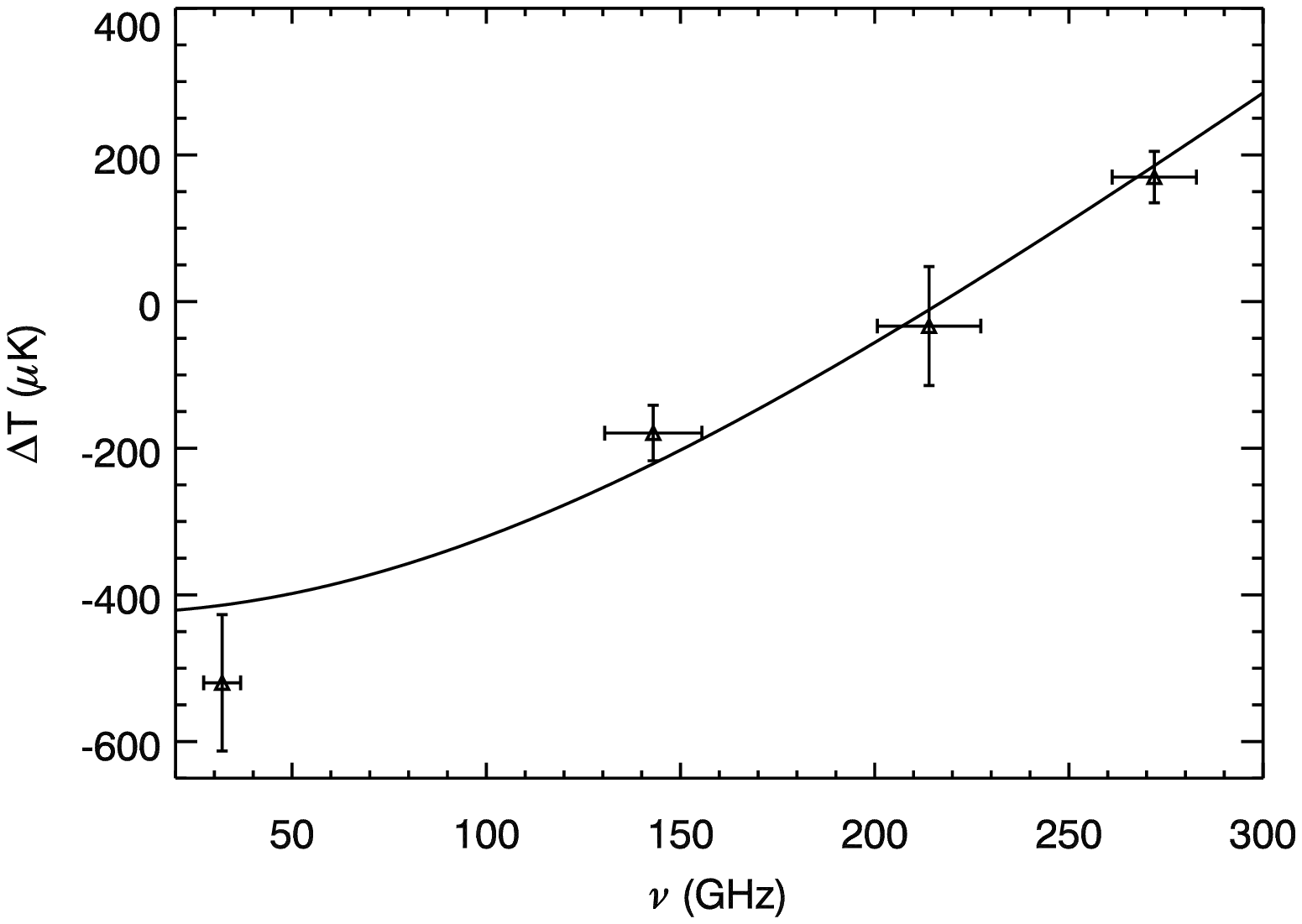}}}\medskip
{\small F{\scriptsize IG}.~2. SZ spectrum of the Coma cluster. The
solid line shows the best fit spectrum (taking isothermal gas with
$kT = 8.2$ keV) to the combined MITO and OVRO (Mason \ea 2001)
measurements, corresponding to $\tau \simeq (4.9 \pm 0.7) \times
10^{-3}$.\label{fig:data}}
\medskip

The work reported here is only the first stage of an extensive
project to upgrade MITO to a bolometer array consisting of 36
elements operating at four frequency bands and with beamsizes down
to 4$'$. Work on the upgraded system is underway, with first
observations scheduled for early 2003. We plan to measure the
effect in a large sample of nearby clusters, with the goal of
determining $H_0$ and cluster masses. Based on the results from
our long campaign to measure the effect in Coma, we are reasonably
confident that this project is quite feasible.

\acknowledgments Contributions to the success of the MITO program
were made by many scientists; in particular, we are grateful to:
S. Masi, who designed and built the cryogenic He-3 system; P.
Lubin, who kindly supplied a low noise cryogenic FET; T. Encrenaz,
for her FIR spectra of planets, and to P. de Bernardis for various
useful suggestions. This work has been supported by CNR (Testa
Grigia Laboratory is a facility of Sezione IFSI in Turin)
COFIN-MIUR 1998, \& 2000, by ASI contract BAR, and by a NATO
Grant.

\end{document}